\documentclass[letterpaper,aps,prd,twocolumn,showpacs,preprintnumbers,nofootinbib]{revtex4}
\usepackage{graphicx}
\usepackage[intlimits,centertags]{amsmath}
\usepackage{amssymb,amsfonts}
\usepackage{gensymb}
\usepackage{mathrsfs}
\usepackage{hyperref}
\hypersetup{pdfauthor=Markus Ahlers and Francis Halzen,pdftitle=Pinpointing Extragalactic Neutrino Sources in Light of Recent IceCube Observations,colorlinks=true,linkcolor=blue,citecolor=blue,filecolor=blue,urlcolor=blue}
\begin{document}

\title{Pinpointing Extragalactic Neutrino Sources in Light of Recent IceCube Observations}

\author{Markus~Ahlers} 
\affiliation{Wisconsin IceCube Particle Astrophysics Center (WIPAC) and Department of Physics,\\ University of Wisconsin, Madison, WI 53706, USA}

\author{Francis~Halzen} 
\affiliation{Wisconsin IceCube Particle Astrophysics Center (WIPAC) and Department of Physics,\\ University of Wisconsin, Madison, WI 53706, USA}

\begin{abstract}
The IceCube Collaboration has recently reported the observation of a flux of high-energy astrophysical neutrinos. The angular distribution of events is consistent with an isotropic arrival direction of neutrinos which is expected for an extragalactic origin. We estimate the prospects of detecting individual neutrino sources from a quasi-diffuse superposition of many extragalactic sources at the level of the IceCube observation. Our analysis takes into account ensemble variations of the source distribution as well as the event statistics of individual sources. We show that IceCube in its present configuration is sensitive to rare $\lesssim10^{-8}\,{\rm Mpc}^{-3}\,{\rm yr}^{-1}$ transient source classes within 5 years of operation via the observation of multiplets. Identification of time-independent sources is more challenging due to larger backgrounds. We estimate that during the same period IceCube is sensitive to sparse sources with densities of $\lesssim10^{-6}\,{\rm Mpc}^{-3}$ via association of events with the closest 100 sources of an ensemble. We show that a next-generation neutrino observatory with 5 times the effective area of IceCube and otherwise similar detector performance would increase the sensitivity to source densities and rates by about two orders of magnitude.
\end{abstract}

\pacs{98.70.Sa,95.55.Vj}

\maketitle

\section{Introduction}\label{sec1}

The high-energy cosmic ray, $\gamma$-ray and neutrino emission of the Universe are fascinating phenomena based on poorly understood non-thermal processes in astrophysical environments. Whereas high-energy $\gamma$-rays can be produced via leptonic processes like inverse Compton scattering or bremsstrahlung, the production of neutrinos require interactions of high energy cosmic rays with radiation or matter. Hence, cosmic neutrinos are the unambiguous tracers of cosmic ray interactions in our Universe and neutrino astronomy promises to unravel their sources.

High energy astrophysical neutrinos become visible in neutrino observatories once they interact in the detector vicinity via charged and neutral current interactions. The high energy secondary particles produced in these interactions are detected in transparent media like water or ice via Cherenkov light emission. The cross sections of these processes are very low. For instance, at PeV energies the neutrino interacts with nucleons with a cross section of about 1 nbarn. Hence, for a detector density of the order of $N_A\,{\rm cm}^{-3}$, we expect only a fraction of about $10^{-5}$ to interact in 1 km of the medium. These low event rates and the large background from atmospheric cosmic ray interactions are the experimental challenges for neutrino astronomy.

The IceCube Collaboration~\cite{Aartsen:2013jdh,Aartsen:2014bea} has recently reported the detection of a cosmic flux of high energy neutrinos with a significance of $5.7\sigma$. The flux is consistent with an $E^{-\gamma}$ power spectrum with spectral index $\gamma\simeq2.3\pm0.3$ and equal distribution between flavor and isotropic arrival direction of neutrinos. The best fit $E^{-2}$-flux is given as
\begin{equation}\label{eq:ICnu}
E_\nu^2J^{\rm IC}_\nu(E_\nu) \simeq (0.95\pm0.3)\times10^{-8}\,\frac{{\rm GeV}}{{\rm cm}^{2}\,{\rm s}\,{\rm sr}}
\end{equation}
The origin of these neutrinos is unknown. A statistically weak cluster of events near the Galactic Centre has motivated speculations about a Galactic origin of the signal. These scenarios include the diffuse neutrino emission of Galactic CRs~\cite{Ahlers:2013xia,Kachelriess:2014oma}, the joint emission of Galactic PeVatrons~\cite{Fox:2013oza,Gonzalez-Garcia:2013iha} or extended Galactic structures like the Fermi Bubbles~\cite{Razzaque:2013uoa,Ahlers:2013xia,Lunardini:2013gva} or the Galactic Halo~\cite{Taylor:2014hya}. A possible association with the sub-TeV diffuse Galactic $\gamma$-ray emission~\cite{Neronov:2013lza} and constraints from the non-observation from diffuse Galactic PeV $\gamma$-rays~\cite{Gupta:2013xfa,Ahlers:2013xia} have also been mentioned. More radical suggestions include PeV dark matter decay scenarios~\cite{Feldstein:2013kka,Esmaili:2013gha,Bai:2013nga}. The extension of the events to large Galactic latitudes and the absence of significant event clusters suggest that a significant contribution of the signal originates in extragalactic sources. Possible source candidates include galaxies with intense star-formation~\cite{Loeb:2006tw,Murase:2013rfa,He:2013cqa}, cores of active galactic nuclei~\cite{Stecker1991,Stecker:2013fxa}, low-power $\gamma$-ray bursts~\cite{Waxman:1997ti,Murase:2013ffa}, intergalactic shocks and active galaxies embedded in structured regions~\cite{Berezinsky:1996wx,Murase:2013rfa}. 

The search for transient and continuous neutrino sources has so far been unsuccessful providing upper neutrino flux limits on individual Galactic and extragalactic source candidates~\cite{IceCube:2011ai,AdrianMartinez:2012rp,Abbasi:2012zw,Aartsen:2013uuv,Adrian-Martinez:2014wzf,Adrian-Martinez:2013dsk}. If the IceCube observation is a superposition of individual (possibly extended) sources, in contrast to a truely diffuse emission, these searches should eventually uncover individual sources or at least place upper limits on possible source candidates. However, it is important to keep in mind that the interaction rate of a neutrino is so low that it travels basically unattenuated through the tenuous matter and radiation backgrounds of the Universe over cosmic distances. The un-resolved extragalactic (or quasi-diffuse) flux is hence a superposition of many sources ranging from relatively recent and local objects to old and distant sources as far as the Hubble horizon. This makes the identification of individual point-sources contributing to the IceCube flux challenging. 

In this paper we investigate the necessary performance of a neutrino observatory for the detection of neutrino point-sources in the form of event clusters and in association with close-by sources or source catalogues. We will start in Section~\ref{sec2} with a derivation of the contribution of individual neutrino sources to a quasi-diffuse flux. In Section~\ref{sec3} we derive simple estimates of the required event numbers and rates of the quasi-diffuse emission for the identification of sources assuming a spatially homogeneous local distribution. For continuously emitting neutrino sources an important aspect of the analysis is the background of atmospheric neutrinos and the quasi-diffuse neutrino signal itself. In Section~\ref{sec4} we include these contributions quantitatively via a significance tests introducing simple test statistics for event clusters and source associations. We compare these results with specific source scenarios in Section~\ref{sec5} and conclude in Section~\ref{sec6}. In the following we work in Heaviside-Lorentz units and make use of the abbreviation $A_x = A/(10^xu)$, where $u$ is the (canonical) unit of the quantity $A$. 

\section{Neutrino Point-Sources}\label{sec2}

High-energy neutrinos are produced by the decay of charged pions from hadronic interactions of CRs with radiation ($p\gamma$) and matter ($pp$). The same mechanism produces also high-energy $\gamma$-rays from the formation of neutral pions. On production, the emission rates of neutrinos (summed over neutrinos and anti-neutrinos with flavor $\nu_\alpha$) and $\gamma$-rays are related to the emission rate of CR nucleons ($N$) as
\begin{equation}\label{eq:MM}
\frac{1}{3}\sum_{\alpha}E_\nu Q_{\nu_\alpha} \simeq \frac{K_\pi}{2}  E_\gamma Q_\gamma\simeq \frac{f_\pi}{\kappa}\frac{K_\pi}{1+K_\pi}E_NQ_N\,,
\end{equation}
where the (average) energies are related as $E_\nu\simeq E_\gamma/2\simeq E_N/20$. We will assume in the following that energy losses of the charged pions and muons prior to their decay are unimportant. In this case the relation (\ref{eq:MM}) has to be considered as lower bound on the total CR energy of the sources. The parameter $K_\pi\simeq1$ ($K_\pi\simeq2$) denotes the average ratio between charged and neutral pions in $p\gamma$ ($pp$) interactions and the pion production efficiency, that is related to the optical depth $\tau$ for hadronic interactions with inelasticity $\kappa$ as $f_\pi \simeq 1-\exp(-\kappa\tau)$.

The IceCube flux between 60~TeV to 2~PeV corresponds to $\gamma$-rays between 100~TeV and 4~PeV or CR nucleons between 1~PeV and 40~PeV. Cosmic rays at these energies are deflected in cosmic magnetic fields and can only be used for point-source associations at very high energies. For instance, ultra-high energy (UHE) CR protons beyond 60~EeV are expected to originate within about 200~Mpc due to the strong absorption in the cosmic microwave background (CMB). Even at these extreme energies the deflection via turbulent Galactic magnetic field can be of the order of $0.5^\circ$~\cite{Giacinti:2011uj}. The contribution of extragalactic magnetic fields is more uncertain and can in principle be larger~\cite{Dolag:2004kp}.

Gamma-rays with PeV energies experience no electro-magnetic deflections, but have a short absorption length of about 10~kpc due to $e^+e^-$ production via scattering off of the photons of the CMB. At about 100~TeV interactions with the extragalactic background light limit the propagation distance to Mpc scales. Direct observation of $\gamma$-ray emission in association with the IceCube flux is hence not feasible, unless there is a significant Galactic contribution~\cite{Ahlers:2013xia}. However, the sub-TeV extension of the IceCube signal (expected for a $pp$ origin of the signal) as well as the sub-TeV contribution of cascaded $\gamma$-rays via inverse-Compton scattering of the high-energy $e^+e^-$ can be visible as an (extended) point-source TeV $\gamma$-ray emission and provide an additional constraint for the contribution of close-by sources~\cite{Murase:2013rfa}.

Neutrinos on the other hand have negligible interactions during propagation and are ideal point-source messengers.
The (quasi-)diffuse flux of neutrinos $J$ (in units of ${\rm GeV}^{-1} {\rm s}^{-1} {\rm cm}^{-2} {\rm s}^{-1}$) originating in multiple cosmic sources is simply given by
\begin{equation}
J_\nu(E_\nu) = \frac{1}{4\pi}\int_0^\infty\frac{{\rm d}z}{H(z)}\mathcal{L}_\nu(z,(1+z)E_\nu)\,,
\end{equation}
where $H$ is the red-shift dependent Hubble expansion rate and $\mathcal{L}$ is the spectral emission rate density. In the case of a continuous neutrino emission we can decompose this into $\mathcal{L}(z,E) = \mathcal{H}(z)Q_\nu(E)$ where $\mathcal{H}$ is the source density and $Q_\nu$ is the emission rate per source. In the following we will assume evolution following the star-formation rate (SFR)~\cite{Hopkins:2006bw,Yuksel:2008cu} $\mathcal{H}_{\rm SFR}(z) \propto (1+z)^{n_i}$ with $n_i=3.4$ for $z<1$,  $n_i=-0.3$ for $1<z<4$ and  $n_i=-3.5$ otherwise. The red-shift dependence of the source distribution can be parametrized by the energy dependent quantity
\begin{equation}
\xi_z(E) = \int_0^\infty{\rm d}z\frac{H_0}{H(z)}\frac{\mathcal{L}_\nu(z,(1+z)E)}{\mathcal{L}_\nu(0,E)}\,.
\end{equation}
For the special case of power-law spectra $\mathcal{L}_\nu(E)\propto E^{-\gamma}$ this quantity is energy independent and we will assume the case $\gamma=2$ in the following corresponding to the fit (\ref{eq:ICnu}). In this case we have $\xi_z\simeq2.4$ assuming evolution with SFR of the CR sources. For a source distribution with no evolution in the local ($z<2$) Universe this reduces to $\xi_z\simeq0.5$. 

From this we can estimate the contribution of an individual continuously emitting point sources. For a distance $d = d_1 10$~Mpc the mean neutrino flux is given as
\begin{align}\label{eq:JPS}
E_\nu^2J_\nu 
\simeq \frac{(0.9\pm0.3)\times10^{-12}}{\xi_{z, 2.4}\mathcal{H}_{0, -5}d_1^{2}} \frac{\rm TeV}{{\rm cm}^{2}\,{\rm s}}\,,
\end{align}
where $\mathcal{H}_0 = \mathcal{H}_{0, -5} 10^{-5} {\rm Mpc}^{-3}$ is the local source density. For this choice of parameters the contribution of a source is consistent with upper limits of neutrino point sources in an un-binned search~\cite{Aartsen:2013uuv}. In the case of transient sources we write instead $\mathcal{L}(z,E) = {\dot{\mathcal{H}}}(z){\rm d}N/{\rm d}E(E)$ with transient rate density ${\dot{\mathcal{H}}}(z)$ and spectrum ${\rm d}N/{\rm d}E$ of an individual flare. In this case the mean neutrino fluence $F$ from an individual transient can be expressed as 
\begin{align}\label{eq:FPS}
E_\nu^2F_\nu 
\simeq \frac{0.3\pm0.1}{\xi_{z, 2.4}{\dot{\mathcal{H}}}_{0, -6}d_1^{2}} \frac{\rm GeV}{{\rm cm}^{2}}\,,
\end{align}
where ${\dot{\mathcal{H}}}_0 = {\dot{\mathcal{H}}}_{0, -6}10^{-6} {\rm Mpc}^{-3}{\rm yr}^{-1}$ is the local flaring/burst density rate.
The sensitivity of IceCube for triggered transient sources lies at about $0.1{\rm GeV}/{\rm cm}^2$ depending on zenith angle and emission time scale~\cite{IceCube:2011ai,Abbasi:2012zw}.

The previous estimates depend on the distance of the source. In the case of a large number of sources we can express the probability that the closest source contributes with an expected number of events $n$ as \begin{equation}\label{eq:ploc}
p_1(n) \simeq \frac{3}{2}\frac{1}{ n}\left(\frac{ n(\hat{r})}{ n}\right)^\frac{3}{2}e^{-\left(\frac{ n(\hat{r})}{ n}\right)^\frac{3}{2}}
\end{equation}
where the distance $\hat{r}$ is defined via $\mathcal{H}_0\hat{r}^3 \Delta \Omega/3=1$, {\it i.e.}~the radius of a sphere where we expect one source in the experimental field of view (FoV) $\Delta \Omega$. The general probability distribution of the $k$th-closest source is given in Appendix~\ref{app1}.

\begin{figure}[t]\centering
\includegraphics[width=\linewidth]{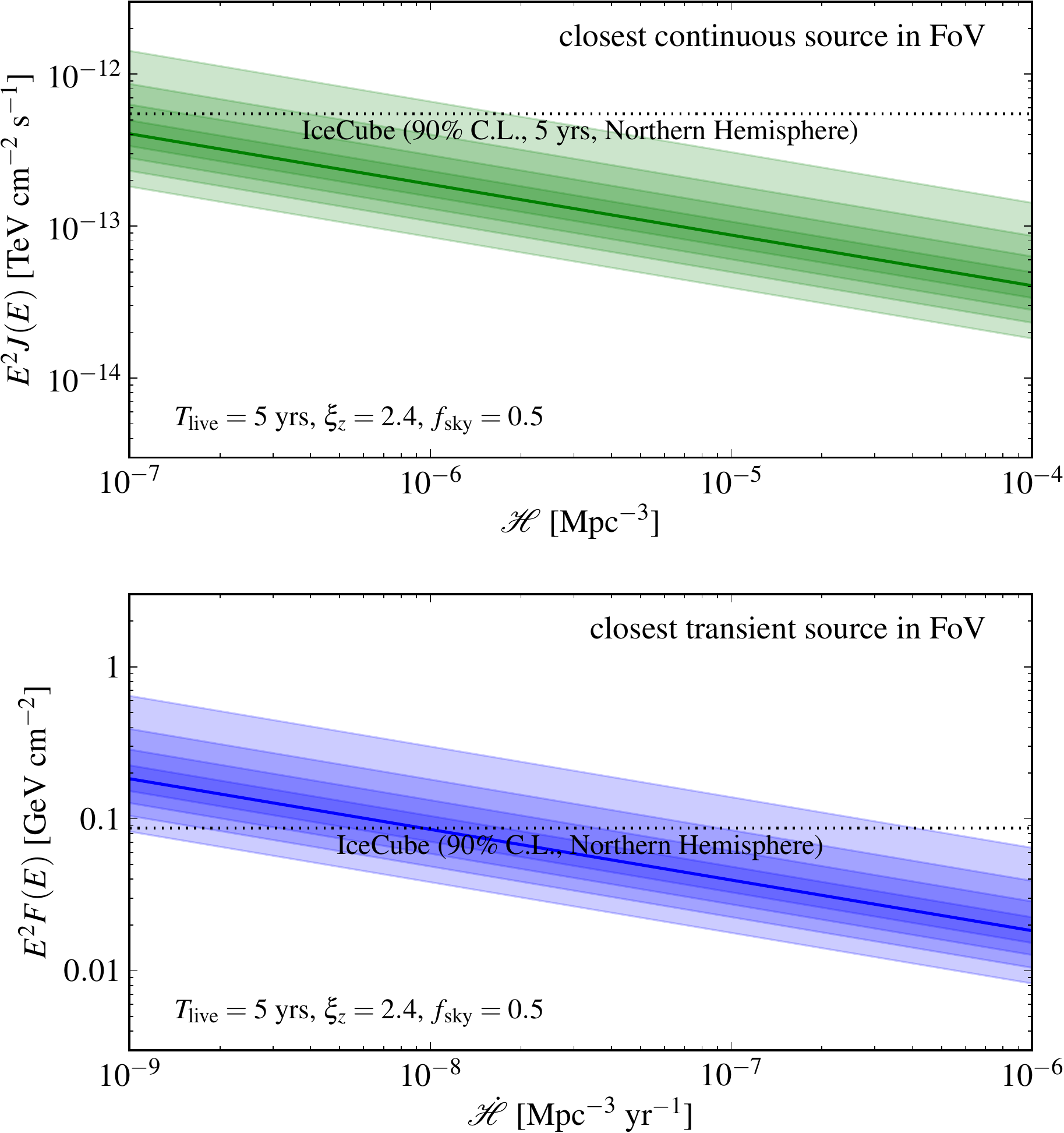}
\caption[]{Point sources sensitivity for continuous (top) and transient (bottom) sources. The shaded areas show the $10\%$ quantiles around the median expectation from the closest sources of the ensemble. The dotted horizontal line show the IceCube sensitivity after five years estimated for a muon energy threshold of 10~TeV (see main text).}\label{fig1}
\end{figure}

In Figure~\ref{fig1} we show the contribution of the closest continuous or transient source in terms of the density of the underlying source population and as $10\%$ quantiles around the median (solid lines) according to Eq.~(\ref{eq:ploc}) and Eqs.~(\ref{eq:JPS}) and (\ref{eq:FPS}), respectively. We assume an observation time $T_{\rm live}=5$~yrs for the total number of transient sources. The dotted horizontal lines show the estimated sensitivity of IceCube to continuous and transient sources for a detector live-time of five years assuming an event rate of $\dot{N}\sim 50~{\rm yr}^{-1}$ above a muon energy threshold of 10~TeV and low background~\cite{Karle2014}. The estimated sensitivity (90\% C.L.) shown in Fig.~\ref{fig1} is then simply $J \simeq f_{\rm sky}4\pi J^{\rm IC}_\nu\times 2.3/(T_{\rm live}\dot{N})$ for event associations with continuous sources and $F \simeq f_{\rm sky}4\pi  J^{\rm IC}_\nu\times 2.3/\dot{N}$ for events triggered by transient sources, where we introduced the fractional sky coverage $f_{\rm sky}=\Delta\Omega/(4\pi)$. We will provide a more precise estimate of the detector sensitivity including backgrounds in the following sections.

These results already indicate that the non-observation of individual neutrino sources (and in particular the closest one) is consistent with the hypothesis of an extragalactic origin of the recent IceCube observation (\ref{eq:ICnu}) for sufficiently large source densities and/or rates. On the other hand, the identification of individual neutrino sources with the continued observation with IceCube over the next years will be challenging unless the source distribution is sufficiently sparse and/or rare. In the next sections we will make this statement more quantitative and discuss the required event numbers and search strategies for an identification of the sources with IceCube or next-generation neutrino observatories.

\section{Point Source Statistics}\label{sec3}

A model-independent identification of neutrino sources can be the detection of spatial or temporal clusters with total number of $m$ neutrino events or more. The required value for $m$ needed for a significant detection depends on the density of sources as well as the expected number of signal and background events. The neutrino event clusters will be most likely associated with local neutrino sources and we can hence simplify the discussion by considering Euclidean space, where we neglect redshift scaling of energy and comoving volume. The contribution from a single local source at distance $r\leq H_0^{-1}$ can be expressed as
\begin{equation}\label{eq:nr}
 n(r) 
\simeq \frac{H_0}{f_{\rm sky}4\pi r^2\xi_z}\times\begin{cases} N/\mathcal{H}_0& \text{(continuous)}\\ \dot{ N}/{\dot{\mathcal{H}}}_0& \text{(transient)}\end{cases}
\end{equation}
In the following we will derive results in terms of the expected number of events $ N$ or $\dot{ N} =  N/T$ of all neutrino sources. 

As a back-of-the-envelope estimate of the required total event numbers for the observation of event multiplets we can consider the contribution of the closest source of the ensemble. For a fractional sky coverage $f_{\rm sky}$ and local source density $\mathcal{H}_0$ we expect one source in the FoV within a sphere of volume $V_1 = 1/(f_{\rm sky}\mathcal{H}_0)$. The total number of events that we expect from this volume is given by the integral of Eq.~(\ref{eq:nr}) over $V_1$ and yields $m =  N(V_1/V_H)^\frac{1}{3}/\xi_z$ where we introduce the Hubble volume $V_H = 4\pi/(3H_0^3)$. Note, that $V_H/V_1$ correspond to the effective number of sources in the FoV. In the case of continuous sources we arrive then at an expected total event number for $m$ local events of\begin{equation}\label{eq:Ncont}
 N_{\rm cont} \simeq 740\,\left(\frac{m}{2}\right)\,\xi_{z, 2.4}\,\left(f_{\rm sky}\,\mathcal{H}_{0,-5}\right)^{\frac{1}{3}}\,.
\end{equation}
In the case of transient sources we have to take into account that the number of sources is increasing with observation time $T_{\rm live} =  N/\dot{ N}$. Solving in terms of the total observation rate $\dot{ N}$ we arrive at 
\begin{equation}\label{eq:Ntrans}
 N_{\rm trans} \simeq 637\left(\frac{m}{2}\right)^\frac{3}{2}\,\xi_{z, 2.4}^\frac{3}{2}\,\left(f_{\rm sky}{{\dot{\mathcal{H}}}_{0,-6}}/{\dot{ N}_2}\right)^\frac{1}{2}\,,
\end{equation}
with an event rate $\dot{ N} = 100\dot{ N}_2/{\rm yr}$. Note, that the event clusters in the transient case should also show a strong temporal coincidence within the burst or flaring time-scale of the source. This fact is important for the comparison with background events due to random clusters of atmospheric neutrinos and muons.

Instead of searching for auto-correlations of events we can also look for cross-correlations with catalogues of candidate sources. For simplicity, let's assume that the catalogue is locally complete up to a distance $r_{\rm cat}$ containing $C = f_{\rm sky}\mathcal{H}_0V_{\rm cat}$ local neutrino sources in the FoV. Following the same line of arguments as in the case of event clustering we can estimate that $m =  N(V_{\rm cat}/V_H)^\frac{1}{3}/\xi_z$ events are expected to correlate with sources of the catalogue. Note that this expression is only valid for $V_{\rm cat}<V_H$; for complete catalogues we simply have $N(m) = m$. For transient sources we assume that the catalogue itself grows in time and remains complete within a distance $r_{\rm cat}$. In this case the expected total number of events for $m$ associations with sources of the catalogue is
\begin{equation}\label{eq:Nass}
 N_{\rm ass} \simeq 107\, m\,\xi_{z, 2.4}r_{\rm cat, 2}^{-1}f_{\rm sky}^{-\frac{1}{3}}\,.
\end{equation}
The result is the same for continuous and transient sources if we assume that the catalogue grows in time for transient sources. In the case of extended source catalogues with $r_{\rm cat}\sim 1/H_0$ already a few signal events can be sufficient for a point-source association. This requires sources with powerful electromagnetic emission which is typical for transient sources like GRBs or AGN flares. 

A crucial aspect for a statistically significant detection of local neutrino sources are the backgrounds of distant neutrino sources as well as atmospheric showers. For transient sources the time-stamp of the signal can lead to a significant reduction of backgrounds. In this case Eqs.~(\ref{eq:Ntrans}) and (\ref{eq:Nass}) already indicate the number of events required for source identifications. However, continuous sources require a more careful statistical discussion. The previous estimates suggest that the required number of observed signal events $N$ has to reach at least a level of 100 before we can expect to observe an association of events with local continuous sources. 

The classical muon neutrino search has a sky coverage of about $f_{\rm sky}\lesssim 0.5$. For the following analysis we will assume that a suitable application of low level event filters has already reduced the background to a level $S/B$. The significance of spatial clustering or association of events depends on the experimental angular resolution $\Delta\theta$. This defines the effective number of bins in the sky as $n_{\rm bin} \simeq 2f_{\rm sky}/(1-\cos(\Delta\theta))$. For instance, we have $n_{\rm bin}\simeq6600$ for $\Delta\theta\simeq 1^\circ$ and $f_{\rm sky}=0.5$, but only $n_{\rm bin}\simeq66$ for $\Delta\theta\simeq 10^\circ$. The expected number of evens from a random distribution of $N_{\rm bg}$ background events is $m\simeq {N}_{\rm bg}/n_{\rm bin}$ per bin. Hence, for poorly resolved cascade events with $\gtrsim10^\circ$ resolution even a low background contribution of $S/B\simeq 1$ already produces random event clusters and association.

In the case of transient sources with burst or flaring time window $\Delta T\ll T_{\rm live}$ we can make use of the fact that the CR background is continuous. We can account for this by redefining the effective number of bins as $n_{\rm bin}\to T_{\rm live}/\Delta Tn_{\rm bin}$. In general, $T_{\rm live}/\Delta T$ is expected to be very large and depends on the specific source. The background cluster probability in the transient case is expected to be very low. 

\section{Significance Test}\label{sec4}

In order to test the statistical significance of neutrino point-sources we introduce two simple test statistics (TS), {\it i)}  ${\rm TS}_1 = \max\lbrace {\bf k}\rbrace$ for cluster tests and {\it ii)} ${\rm TS}_2 = \sum_{N_s} k_i$ for a source association with $N_s$ closest sources. If ${\bf k}$ is the experimental result and ${\bf k}_0$ a possible background distribution we define the ensemble-averaged $p$-value as
\begin{equation}
p= \int\prod_{i=1}^{N_s}{\rm d}n_i\,p_{\rm PS}(n_i)\!\!\!\!\!\!\!\!\!\!\!\!\sum_{{\rm TS}({\bf k}_0)\geq{\rm TS}({\bf k})}\!\!\!\!\!\!\!\!\!\!\!\!P({\bf k})P_0({\bf k}_0)\,,
\end{equation}
where $p_{\rm PS}(n)$ is the probability (\ref{eq:pPS}) for an individual source out of $N_s$ sources to contribute with an expectation value $n$. The event probability distributions for signal $P$ and background $P_0$ are products of Poisson distributions for the content of the individual bins (see Appendix~\ref{app1}). 

One advantage of these test statistics is that the $p$-value reduces to simple analytic formulae in the background-free case, which is typically the case for transient sources. In this case we simply have $p({\rm TS}_1) \simeq 1-P_{\rm cl}(2)$ with Eq.~(\ref{eq:Pcl}) and $p({\rm TS}_2) \simeq 1-P_{\rm ass}(1)$ with Eq.~(\ref{eq:Pass}) derived in Appendix~\ref{app2}. In Appendix~\ref{app2} we also show that in the background-free case the required total signal event numbers at the significance level $p=0.1$ are the same as in Eq.~(\ref{eq:Ncont}) or (\ref{eq:Ntrans}) with the replacement $m\to M\simeq 3.57$. From this we can estimate that IceCube is sensitive to multiplets of rare transient sources $\dot{\mathcal{H}}_0\lesssim10^{-8}\,{\rm Mpc}^{-3}\,{\rm yr}^{-1}$ after five years of observation.

In the case of continuous sources the background probability is not negligible. In the following we discuss two experimental scenarios parametrized by the angular resolution $\Delta\theta$, partial sky fraction of the FoV $f_{\rm sky}$ and the signal-to-background ratio $S/B$ of the observation. In the present IceCube 86-string configuration we expect about 40-60 signal events per year in the Northern Hemisphere above a muon energy threshold of 10~TeV, depending on the spectral index of the observation~\cite{Karle2014}. The atmospheric muon neutrino background is at the level of about 500 events per year. The angular resolution of these events at about 10~TeV is of the order of $0.6^\circ$. We will hence use the combination {\it a)} $f_{\rm sky}=0.5$, $S/B = 0.1$ and $\Delta\theta=0.6^\circ$ as a first detector performance benchmark. At muon energies above about 100 TeV the signal-to-background ratio in the Northern Hemisphere increases to about $S/B=1$ with an increased angular resolution. However, due to neutrino absorptions inside the Earth the effective FoV is reduced. In this case we use as a second performance benchmark {\it b)} $f_{\rm sky}=0.25$, $S/B = 1$ and $\Delta\theta=0.3^\circ$.

\begin{figure*}[t]\centering
\includegraphics[width=\linewidth]{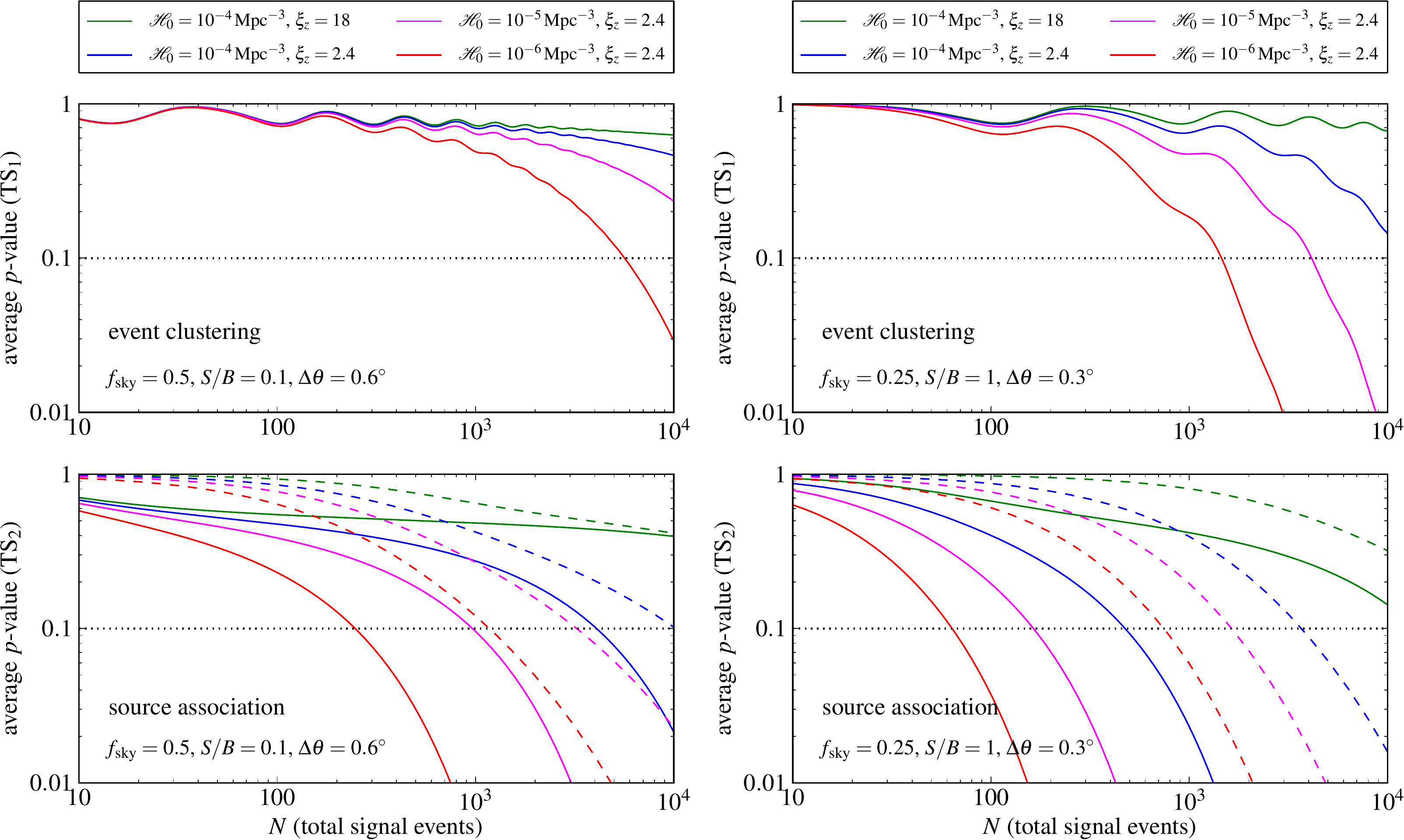}
\caption[]{Results of a significance test of continuous point-sources. 
{\bf Top Panels:} The results for the test statistic ${\rm TS}_1$ looking for significance of event clusters in a map. The four different scenarios are parametrized in terms of an increasing expected total number of signal events $ N$ and the fixed combination of local source density $\mathcal{H}_0$ and source evolution $\xi_z$. The dashed horizontal line indicates the sensitivity threshold of $1-p=90\%$. We show results for a benchmark detector performance {\it a)} $S/B=0.1$, $f_{\rm sky}=0.5$ and $\Delta\theta =0.6^\circ$ (left panel) and {\it b)} $S/B=1$, $f_{\rm sky}=0.25$ and $\Delta\theta =0.3^\circ$ (right panel), corresponding to muon energy thresholds of about $10$~TeV and $100$~TeV, respectively~\cite{Karle2014}. As a reference value, the signal rate of the full IceCube detector in the Northern Hemisphere is about $\dot N \simeq 50\, {\rm yr}^{-1}$ above muon energies of 10~TeV and $\dot N \simeq 10\, {\rm yr}^{-1}$ above 100~TeV. 
{\bf Lower Panels:} Same as top panels but now showing the results for the test statistic ${\rm TS}_2$ looking for association with 100 closest members of the ensemble (solid line) and the closest source (dashed line).}\label{fig2}
\end{figure*}

In Figure~\ref{fig2} we show results of the averaged $p$-value for different astrophysical scenarios of continuous neutrino sources and detector parameters. As usual the sensitivity is defined as a 90\% confidence level (C.L.) of the signal hypothesis or, equivalently, $p=0.1$. The upper panel of Fig.~\ref{fig2} shows the results for ${\rm TS}_1$ for the search of event clusters for the benchmark point {\it a)} (left panel) and {\it b)} (right panel) . Even with high $S/B$ and good angular resolution the significance of event clusters requires event numbers of $10^3$ in the most favorable astrophysical scenario of low source densities of $10^{-6}~{\rm Mpc}^{-3}$. Hence, a model-independent identification of extragalactic neutrino sources is challenging even with a detector with ten times the event rates of the present IceCube configuration. Note that the oscillatory pattern in the $p$-values is due to the averaging over ensembles and experimental realizations. For an actual realization with maximum multiplet $k_{\rm max}$ the $p$-value is simply $p=1-P_{\rm bg}(k_{\rm max})$ which is continuously decreasing with $N$. 

The situation for continuous sources improves in the case of associations of neutrino events with candidate source catalogues tested by the test statistic ${\rm TS}_2$. We show results in the bottom panels of Fig.~\ref{fig2} for the benchmark points {\it a)} (left panel) and {\it b)} (right panel), respectively. The dashed lines show the results of association with the closest source of the ensemble following the distribution (\ref{eq:ploc}). The detector performance used in the middle panel corresponds to the case shown in Fig.~\ref{fig1}. For instance, the optimistic source density of $10^{-6}~{\rm Mpc}^{-3}$ requires 1000 event for $p=0.1$. With a present IceCube rate of 50 signal events per year this requires an experimental live-time of $T_{\rm live}\simeq 20$~yrs, consistent with the sensitivity level indicated in Fig.~\ref{fig1}. This provides an {\it a posteriori} justification of the background-free estimate.

The solid lines in the lower panels of Fig.~\ref{fig2} show the results of ${\rm TS}_2$ in the case of an association of events with the 100 closest sources. In the case of the detector performance {\it b)} with a high purity of events of $S/B\simeq1$ and good angular resolution the observation of $50$--$500$ events is required for source densities of $10^{-6}$--$10^{-4}\,{\rm Mpc}^{-3}$ and a moderate source evolution following the star-formation rate ($\xi_z\simeq2.4$). For a detector with  $\dot{N}\simeq 50~{\rm yr}^{-1}$ corresponding to five times IceCube's event rates above muon energies of 100~TeV this would require 1--10 years of observation. 

\section{Source Candidates}\label{sec5}

For a given source density $\mathcal{H}$ or rate density ${\dot{\mathcal{H}}}$ of neutrino sources we can derive a lower limit on the energy budget of the individual sources. The neutrino emission extends in the energy range from $60$~TeV to $2$PeV and hence, conservatively, the corresponding range of the underlying CR population ranges from $E_-\simeq m_p$ to $E_+\simeq 40$~PeV. The mean integrated emission rate of continuous point sources can then be estimated as
\begin{equation}\label{eq:Qnu}
\int\limits_{E_-}^{E_+}{\rm d} EEQ_N 
\simeq8.9\times10^{42}\frac{\mathcal{R}_{17.5}}{f_\pi\xi_{z,2.4}\mathcal{H}_{0, -5}}\frac{\rm erg}{\rm s}\,,
\end{equation}
where we assumed $K_\pi=2$ ($pp$) and an energy independent pion fraction efficiency $f_\pi$ between $E_-$ and $E_+$. We also assume a spectral index $\gamma=2$ corresponding to a bolometric correction factor $\mathcal{R} = \ln(E_+/E_-) = 17.5\mathcal{R}_{17.5}$. In the case of transient sources the emission spectra ${\rm d}N/{\rm d}E$ of neutrinos, $\gamma$-rays and CRs are related via the analogue of Eq.~(\ref{eq:MM}). The mean emission of an individual point source corresponding to the IceCube observation can then be estimated as
\begin{equation}\label{eq:Nnu}
\int\limits_{E_-}^{E_+}{\rm d}EE\frac{{\rm d}N_N}{{\rm d}E}
\simeq2.8\times10^{51}\frac{\mathcal{R}_{17.5}}{f_\pi\xi_{z,2.4}{\dot{\mathcal{H}}}_{0, -6}}{\rm erg}\,.
\end{equation}
Note, that for UHE CR sources reaching energies of the order of $E_+\simeq10^{12}$~GeV we have $\mathcal{R} \simeq 27.6$. We will discuss in the following specific source scenarios in terms of the required source energy budget, density, rate and evolution.

\subsubsection{Active Galactic Nuclei}

The population of hard x-ray emitting AGN has a peak luminosity at about $10^{43}$--$10^{44}$~erg/s with a source density of $\mathcal{H}_{0}\simeq 10^{-5}$--$10^{-4}{\rm Mpc}^{-3}$\cite{Barger:2004is,Tueller:2007rk}. If we assume that CRs are accelerated to a comparable power we can see that the requirements of Eq.~(\ref{eq:Qnu}) are fulfilled with high pion production efficiencies. We assume that the neutrino luminosity is proportional to the x-ray luminosity and follow the model of~\cite{Barger:2004is,Aird:2009sg}. In this case the red-shift evolution factor is given by $\xi_z \simeq 3.6$ and $\mathcal{H}_0\simeq 10^{-5}{\rm Mpc}^{-3}$. 

For instance, the close-by radio galaxy Cen A at distance of about 4~Mpc has presently an upper limit for muon neutrinos that is about 40 times higher then Eq.~(\ref{eq:JPS})~\cite{Aartsen:2013uuv}. However, note that the TeV $\gamma$-ray emission is about $2.5\times10^{-13}{\rm TeV}^{-1}{\rm cm}^{-2}{\rm s}^{-1}$~\cite{Aharonian:2009xn}, which is almost two orders of magnitude lower than the hadronic $\gamma$-ray emission expected from the relations (\ref{eq:MM}). Another close radio galaxy M87 at 16~Mpc has a stronger upper limit on continuous neutrino emission that is only a factor 5 higher than the prediction (\ref{eq:JPS}). The observed TeV $\gamma$-ray emission is of the order of $6\times10^{-12}{\rm TeV}^{-1}{\rm cm}^{-2}{\rm s}^{-1}$~\cite{Aleksic:2012uma}, which is one order of magnitude lower than the estimate (\ref{eq:MM}).

In the unified AGN model the population of blazars is a fraction of those radio-load AGNs where the jet emission is aligned with the observation axis. For instance, the isotropic equivalent density of flat-spectrum radio quasars peaks at a jet luminosity of $10^{47}$~erg/s with $\mathcal{H}_{0}\simeq 10^{-9}{\rm Mpc}^{-3}$~\cite{Ajello:2011zi}. We assume the same evolution as the underlying AGN source population giving $\xi_z\simeq3.6$. Two of the closest blazars are Mrk 421 ($\sim$130~Mpc) and Mrk 501 ($\sim$140~Mpc). In this case the neutrino upper limits are one order of magnitude stronger than the average emission predicted by Eq.~(\ref{eq:JPS}) and disfavor a blazar origin of the IceCube observation.

Giant AGN flares with an energy release of the order of $10^{51}$~erg have been speculated as a possible source of UHE CRs~\cite{Farrar:2008ex}. Their event rate is estimated to be of the order of ${\dot{\mathcal{H}}}_{0}\simeq 10^{-6}~{\rm Mpc}^{-3}{\rm yr}^{-1}$ consistent with Eq.~(\ref{eq:Nnu}) for high pion production efficiencies. Note that the present IceCube configuration would only be sensitive to these flares via triggering on known close-by sources. On the other hand a detector with five times the effective area would be sensitive to this source class model-independently via a significant spatial and temporal clustering of events. 

\subsubsection{Starburst Galaxies}

Starburst galaxies show a high star formation rate of $10^{-3} M_\odot {\rm Mpc}^{-3} {\rm yr}^{-1}$. Assuming that CRs are accelerated in supernova shocks with a total CR energy of $10^{50}$~erg at a rate of about $0.1\,{\rm yr}^{-1}$~\cite{Lacki:2013ata} we can estimate a total power of $3\times10^{41}{\rm erg}/{\rm s}$ per source. The redshift evolution of starburst galaxies follows the average SFR at high redshift ($z\gtrsim1$). However, the local contributions is dominated by normal galaxies and starburst only contribute at a level of 10\%. We account for relative evolution of the starburst density by a factor ${\rm min}(0.1+0.9z,1)/0.1$ suggested in Ref.~\cite{Thompson:2006np} resulting in an evolution parameter $\xi\simeq18$. Hence, we can estimate the required average CR power of starburst galaxies to match the IceCube flux as $10^{41}/f_\pi$~erg/s and hence $f_\pi\simeq 1$, {\it i.e.}~CR {\it calorimetry}~\cite{Loeb:2006tw}. 

The expected neutrino flux (\ref{eq:JPS}) on starburst galaxy M82 at about 3.5~Mpc is about 40 times lower than present upper limit~\cite{Aartsen:2013uuv}. The starburst NGC 253 at about 2.5~Mpc lies in the the Southern Hemisphere and point-source neutrino limits from IceCube are weaker. Intriguingly, the expected photon point-source flux from Eq.~(\ref{eq:MM}) is comparable to that observed in TeV $\gamma$-ray emission for M82 and NGC 253~\cite{Karlsson:2009hd,Abramowski:2012xy}. This agrees with the result of Ref.~\cite{Lacki:2010vs} based on proton calorimetry.

Note that the large evolution factor $\xi_z\simeq18$ in this scenario produces a large background consisting of distant neutrino emitters. The detection of individual neutrino sources is hence challenging. The corresponding results of the significance tests are shown as green lines in Fig.~\ref{fig2}. For all scenarios the required signal event number exceed $10^4$.
 
\subsubsection{Gamma-Ray Bursts}

Long duration GRBs following the collapse of massive stars occur with an (isotropic equivalent) rate density of ${\dot{\mathcal{H}}}_{0}\simeq 10^{-9} {\rm Mpc}^{-3}\,{\rm yr}^{-1}$~\cite{Wanderman:2009es} and energy of $M_\odot \simeq 2\times10^{54}$~erg. Following the GRB evolution model of Ref.~\cite{Wanderman:2009es} gives an evolution parameter of $\xi_z\simeq1.9$. Again, this is consistent with the required power (\ref{eq:Nnu}) if the pion production efficiency is high. However, in the case of GRBs there are strong bounds on the neutrino emission in coincidence with the $\gamma$-ray display~\cite{Abbasi:2012zw} which are a factor 5 lower than the observed diffuse flux (\ref{eq:ICnu}). 

It has been speculated that low-power GRBs that are unobservable via their burst might be more efficient neutrino factories~\cite{Murase:2013ffa}. In this case the strong IceCube bounds don't apply~\cite{Abbasi:2012zw}. However, even for IceCube's moderate signal event rates of the order of $50~{\rm yr}^{-1}$ above a muon energy threshold of 10~TeV event clusters in time and space are expected to appear already within one year of observation ($m\to M\simeq3.57$ for 90\% C.L.~of doublets in Eq.~(\ref{eq:Ntrans})). Hence, already the present neutrino data can constrain this scenario model-independently. The triggered search on known GRBs is even more sensitive~\cite{Abbasi:2012zw}.

\subsubsection{UHE CR Calorimeters}

The required emission rate density of UHE CR proton sources can be estimated as $E^2_pQ_p(E_p) \simeq (1-2)\times10^{44}\,{\rm erg}\,{\rm Mpc}^{-3}\,{\rm yr}^{-1}$~\cite{Ahlers:2012rz}. If we assume that the emission spectrum $Q_p$ extends to lower energies the corresponding neutrino flux can be determined via Eq.~(\ref{eq:MM}) and can be expressed as
Neglecting a redshift dependence of $f_\pi$ this translates into an $E^{-2}$ flux ($K_\pi=2$) of 
\begin{equation}
E_\nu^2J_\nu(E_\nu) \simeq \xi_{z,2.4} f_\pi(3-6)\times10^{-8}\,\frac{\rm GeV}{{\rm cm}^{2}\,{\rm s}\,{\rm sr}}\,,\label{eq:UHECRnu}
\end{equation}
The similarity of Eqs.~(\ref{eq:ICnu}) and (\ref{eq:UHECRnu}) suggests that the IceCube flux is related to the sources of UHE CRs. The limit $f_\pi\gtrsim1$ in Eq.~(\ref{eq:UHECRnu}) corresponds to the Waxman-Bahcall bound~\cite{Waxman:1998yy,Bahcall:1999yr}. Interestingly, the observation (\ref{eq:ICnu}) is close to this bounds requiring high pion production efficiencies. 

On the other hand, the energy loss due to pion production during acceleration has to be sufficiently low for UHE CR sources in order to compete with the energy gain per acceleration cycle. A natural solution to this fine-tuning problem occurs if the acceleration site is distinct from the CR calorimeter. Such a scenario could be provided in starburst galaxies, where the sub-PeV neutrino production happens during CR propagation in the dusty starburst environment after they have been released from transient UHE CR sources~\cite{Katz:2013ooa}. Other scenarios consider massive galaxy clusters with local densities of the order of $\mathcal{H}_0\simeq 10^{-6}$--$10^{-5}{\rm Mpc}^{-3}$ as CR calorimeters~\cite{Berezinsky:1996wx,Murase:2008yt}. Note that the neutrino emission from CR propagation in CR calorimeters is expected to be continuous even if the sources of UHE CRs are transients, such as flaring AGNs or GRBs. 

\section{Conclusion}\label{sec6}

In this work we have studied the probability of identifying extra-galactic neutrino sources as the likely origin of the astrophysical neutrino flux observed by the IceCube Collaboration. We have derived the expected flux of individual continuous and transient sources based on the IceCube observation and depending on source density and rate. Our analysis takes into account the ensemble variation of the source distribution, their cosmic evolution and the Poisson statistics of weak nearby sources. In order to account for the background of atmospheric neutrinos and distant neutrino sources we introduced test statistics for significance tests of event clusters and source associations.

Our findings are as follows. A model-independent observation of neutrino event clusters from a continuously emitting source population is challenging due to large atmospheric backgrounds unless the experimental angular resolution can be improved significantly. For transient sources these backgrounds are in general much lower due to the correlation of signal events in space and time. We estimate that the present IceCube detector is sensitive to rare $\dot{\mathcal{H}}_0\lesssim10^{-8}~{\rm Mpc}^{-3}\,{\rm yr}^{-1}$ source classes within five years of operation. A next generation telescope with a five times extended effective muon area would be sensitive to multiplets from transient sources with $\dot{\mathcal{H}}_0\lesssim10^{-6}~{\rm Mpc}^{-3}\,{\rm yr}^{-1}$. 

The sensitivity to the underlying source population can be increased in model-dependent associations of events with source catalogues. We estimate that the IceCube detector is already sensitive to sparse continuous sources with $\mathcal{H}_0\lesssim10^{-6}~{\rm Mpc}^{-3}$ as well as transient sources at the level of $\dot{\mathcal{H}}_0\lesssim10^{-5}~{\rm Mpc}^{-3}\,{\rm yr}^{-1}$ via the association of events with the 100 closest sources of the ensemble. Again, a next-generation detector with 5 times the effective area as IceCube and otherwise identical performance in terms of angular resolution and FoV would improve the sensitive of these searches by about two orders of magnitude. A few final remarks are in order. 

{\it i)} In this analysis we have only considered the scenario that the IceCube observation has an extragalactic origin. However, a (partial) Galactic origin of the observation is not yet excluded on statistical grounds. The presence of Galactic contributions will most likely emerge as a large scale anisotropy of the signal which has not been discussed in this work. 

{\it ii)} We have expressed our results via ensemble-averaged $p$-values. The shaded regions shown in Fig.~\ref{fig1} indicate the $\pm40\%$ fluctuation around the median. This {\it cosmic variance} of the closest source can hence increase or decrease the prospects of detecting the closest source, depending on the sign of the fluctuation.

{\it iii)} We assumed that the individual sources of the ensemble can be approximated by a universal (average) luminosity. It is straightforward to include a non-trivial luminosity function or other source nuisance parameters in the definition of the point source distribution (\ref{eq:pPS}) and related quantities. 

{\it iv)} Future neutrino observatories may employ an extended surface veto for atmospheric events that would reduce the contribution of atmospheric neutrinos~\cite{Schonert:2008is,Karle2014}. This can increase the signal-to-background ratio at lower muon energy threshold and improve the detection prospects estimated in this analysis. 

{\it v)} The identification of sources via the association of events with source catalogues is model-dependent and the statistical results need to be corrected by trials factors. However, complementary to the identification of individual neutrino sources the study of spectral properties of the quasi-diffuse neutrino flux will help to narrow down valid astrophysical scenarios and should hence limit the number of trials.

{\it vi)} The identification of extra-galactic neutrino point-sources from a quasi-diffuse flux has also been discussed recently in Refs.~\cite{Fargion:2014mda,Anchordoqui:2014yva}. In this analysis we have carefully studied the effect of ensemble and statistical variations and accounted for backgrounds via a significance test. This makes our estimates more robust and less optimistic than Refs.~\cite{Fargion:2014mda,Anchordoqui:2014yva}.

\begin{acknowledgements}
We would like to thank the IceCube Collaboration for many fruitful discussions, in particular Jake Feintzeig, Gary Hill, Albrecht Karle, Claudio Kopper and Chris Weaver. We acknowledge support by the U.S. National Science Foundation (NSF) under grants OPP-0236449 and PHY-0236449.
\end{acknowledgements}

\begin{appendix}

\section{Source Distribution}\label{app1}

The probability distribution of single source within $r<R$ is $p(r) = 3r^2/{R^3}$. We can express this in terms of the expected events from a point-source using ${\rm d} n/{\rm d}r = - 2 n/r$ as
\begin{equation}\label{eq:pPS}
p_{\rm PS}(n,n(R)) = \Theta(n-n(R))\frac{3}{2}\frac{1}{ n}\left(\frac{ n(R)}{ n}\right)^{3/2}\,,
\end{equation}
with $n(R)$ defined by Eq.~(\ref{eq:nr}). The expected event distribution for the $k$th-closest of the $N_s$ sources of the ensemble is given by
\begin{multline}
p_k(n) = \Theta(n-n(R))\frac{N_s!}{(k-1)!(N_s-k)!}\\\times\frac{3}{2}\frac{1}{n}\left(\frac{n(R)}{n}\right)^{\frac{3k}{2}}\left[1-\left(\frac{n(R)}{n}\right)^{\frac{3}{2}}\right]^{N_s-k}\,.
\end{multline}
For $N_s\gg k$ we can approximate the last term as
\begin{equation}
\left[1-\left(\frac{n(R)}{n}\right)^{\frac{3}{2}}\right]^{N_s-k}
\simeq e^{-N_s\left(\frac{n(R)}{n}\right)^{\frac{3}{2}}}\,,
\end{equation}
and using $(N_s-k) n(R)^{3/2} \simeq n(\hat{r})^{3/2}$ we arrive at
\begin{equation}
p_k(n) \simeq \frac{3}{2(k-1)!}\frac{1}{n}\left(\frac{n(\hat{r})}{n}\right)^{\frac{3k}{2}}e^{-\left(\frac{ n(\hat{r})}{ n}\right)^\frac{3}{2}}\,.
\end{equation}
For $k=1$ this agrees with Eq.~(\ref{eq:ploc}) for the closest source.

\section{Signal Probability}\label{app2}

In general, the probability of the events ${\bf k}$ is given by the Poisson distribution of events in the individual bins with expected signal events $\lambda_i$ and background $\lambda_{\rm bg}$,
\begin{equation}
P({\bf k}) = \prod_{i=1}^{n_{\rm bin}}\frac{(\lambda_i+\lambda_{\rm bg})^{k_i}}{k_i!}e^{-\lambda_{\rm bg}-\lambda_i}\,,
\end{equation}
where $\sum_i (\lambda_i+\lambda_{\rm bg}) = N_{\rm tot}$. The total number of events can be expressed via the expected number of all signal events and the signal-to-background ratio $S/B$ as $N_{\rm tot}= (1+(S/B)^{-1})N$. The background probability is simply
\begin{equation}
P_0({\bf k}) = \prod_{i=1}^{n_{\rm bin}}\frac{\lambda_0^{k_i}}{k_i!}e^{-\lambda_0}\,,
\end{equation}
with $n_{\rm bin}\lambda_0 = N_{\rm tot}$.

In the case of the test statistics ${\rm TS}_{1/2}$ for event clusters and source associations we can derive simple probabilities in the background free case. The probability of observing less than $m$ events from $ n$ expected events can be expressed via incomplete $\Gamma$-functions as
\begin{equation}
\sum_{k=0}^{m-1}  \frac{ n^k}{k!}e^{- n} = \frac{\Gamma(m, n)}{\Gamma(m)}\,.
\end{equation}
Hence, the probability of observing less than $m$ events from a single source within distance $R$ is given as
\begin{equation}
 P_{\rm PS}(m) = \int\limits_{ n(R)}^\infty {\rm d}n\, p_{\rm PS}(n,n(R))\frac{\Gamma(m, n)}{\Gamma(m)}\,.
\end{equation}
This can be expressed as
\begin{equation}
 P_{\rm PS}(m) = \frac{\Gamma\left(m, n(R)\right)-(n(R))^{3/2}\Gamma\left(m-\frac{3}{2}, n(R)\right)}{(m-1)!}\,.
\end{equation}
The probability of observing at least one cluster of $m$ or more neutrinos from a local source is given by the expression $P_{\rm cl} = 1 -  P_{\rm PS}^{N_s}$.
In the limit $n(R)\ll 1$ and for $m\geq2$ this can be approximated as
\begin{equation}\label{eq:Pcl}
P_{\rm cl}\simeq1-\exp\left(-\frac{\Gamma(m-\frac{3}{2})}{3^{\frac{3}{2}}\Gamma(m)}\left(\frac{N}{\xi_z}\right)^\frac{3}{2}\left(\frac{V_1}{V_H}\right)^\frac{1}{2}\right)
\,.
\end{equation}
Hence, this result is independent of the auxiliary distance $R$ as expected. For a significance level $P$ the required total signal event number is the same as in Eqs.~(\ref{eq:Ncont}) or (\ref{eq:Ntrans}) with the replacement
\begin{equation}\label{eq:mM}
m\to M=\left(\frac{3^\frac{3}{2}\Gamma(m)}{\Gamma(m-\frac{3}{2})}\ln\frac{1}{1-P}\right)^\frac{2}{3}\,.
\end{equation}

For the case of local source associations we can write the probability of observing $m$ or more events as
\begin{multline}\label{eq:Pass}
P_{\rm ass}(m) = 1-\int{\rm d}n_Cp_C(n_C)\prod_{i=1}^{C-1}{\rm d}n_i p_{\rm PS}(n_i,n_C)\\\times\frac{\Gamma(m,\sum_{i=1}^Cn_i)}{\Gamma(m)} \,.
\end{multline}
For large number of local sources $C$ we can approximate this by the average number of expected events as
\begin{equation}
P_{\rm ass}(m) \simeq 1-\frac{\Gamma(m,N(V_C/V_H)^\frac{1}{3}/\xi_z)}{\Gamma(m)}\,.
\end{equation}

\end{appendix}

\bibliography{bibliography}

\end{document}